\documentclass[fleqn,twoside]{article}
\usepackage{espcrc2}

\usepackage{graphicx}
\newcommand{\AmS}{{\protect\the\textfont2
  A\kern-.1667em\lower.5ex\hbox{M}\kern-.125emS}}

\title{
Top Mass Measurements from Jets and the Tevatron Top-Quark Mass
}

\author{Andr\'e H.\ Hoang\address{Max-Planck-Institut f\"ur Physik (Werner-Heisenberg-Institut), 
  F\"ohringer Ring 6, 80805 M\"unchen, Germany}\thanks{Talk given at the
  International Workshop on Top Quark Physics, La Biodola, Isola d'Elba, May
  18-24, 2008.},
 Iain W.\ Stewart\address{Center for Theoretical Physics, Massachusetts Institute of
Technology, Cambridge, MA 02139}\thanks{Talk given at the 2nd Workshop on
Theory, Phenomenology and Experiment in Heavy Flavor Physics, Capri, 2008.}
}
       
\begin{document}

\begin{abstract}
  Theoretical issues are discussed for the measurement of the top-mass using
  jets, including perturbative and non-perturbative effects that relate
  experimental observables to the Lagrangian mass, and appropriate choices for
  mass schemes.  Full account for these issues is given for $e^+e^-\to t\bar t$
  using a factorization theorem for event shapes for massive quarks.
  Implications for the Tevatron top-mass measurement are discussed. A
  mass-scheme, the ``MSR-mass'', is introduced which allows for a precise
  description of observables sensitive to scales $R\ll m$, but at the same time
  does not introduce perturbative matching uncertainties in conversion to the $\overline
  {\rm MS}$ mass.  \vspace{1pc}
\end{abstract}

\maketitle
\phantom{x}
\vspace{-1.5cm}

\section{Introduction}

The top-mass is a key parameter in the standard model. For example, it plays an
important role for analyzing electroweak precision constraints, for predicting
rare decays like $B\to X_s\gamma$ and $K_L\to \pi^0\nu\bar\nu$, as well as for
unraveling the Higgs sector in supersymmetric models. The latest measurement
from the Tevatron~\cite{Group:2008nq}, $m_t^{\rm tev}=172.6\pm 0.8(stat) \pm
1.1(syst)\,{\rm GeV}$, is of very high precision, and brings to mind several
theoretical questions. In what mass-scheme is the value quoted? Are measurements
of the mass of a colored particle with a precision better than $\Lambda_{\rm
  QCD}$ possible using jets, or is this an irreducible uncertainty? How do
perturbative and nonperturbative effects modify the relation between the
experimental observable and the underlying Lagrangian mass parameter?  For the
LHC a $\sim 1\,{\rm GeV}$ systematic uncertainty was obtained from preliminary
ATLAS studies~\cite{Borjanovic:2004ce} (where hurdles to go beyond this level
include understanding the jet-energy scale to better than 1\%). At this level,
understanding the answers to the above questions is important.

To achieve a high precision top-mass, measurements exploit kinematic information
by considering leptons + jets, $p p\to t\bar t X \to ( b q\bar q)(\bar b
\ell\bar\nu)$, or the associated dilepton or all-hadronic channels, and use
Monte Carlo (MC) simulations to reconstruct observables sensitive to the
top-quark four vector, 
and hence $m_t^2$. One of the most sensitive observables in this reconstruction
are the invariant masses $M_t^2 = (\sum_{i\in a} p_i^\mu)^2$ and $M_{\bar t}^2 =
(\sum_{i\in b} p_i^\mu)^2$ of the jets and other decay products produced by the
top and antitop respectively. Here the sets $a$ and $b$ depend on the
jet-algorithm and cuts. In such analyses it might appear natural to think of the
reconstructed top-mass as being associated to the pole-mass $m_t^{\rm pole}$,
since it is designed for sensitivity to the on-shell region $p_t^2\simeq
m_t^2$.  However, as we 
will argue in section~5 below, it is not $m_t^{\rm pole}$ that is being
measured by the Tevatron analyses.

The relation of the pole-mass and any other Lagrangian mass-scheme $m_t(R,\mu)$
can be expressed as a perturbative series, $m_t^{\rm pole} = m_t(R,\mu) +\delta
m_t(R,\mu)$ where
\begin{eqnarray}
\label{mRdef}
 \delta m_t(R,\mu) = R \sum_{n=1}^\infty \sum_{k=0}^n a_{nk}
 \Big[\frac{\alpha_s(\mu)}{4\pi}\Big]^n \ln^k\!\Big(\frac{\mu}{R}\Big),
\end{eqnarray}
with $R$ a dimension-1 scale intrinsic to the scheme, and $a_{nk}$ finite
numerical coefficients.  Theoretically the pole-mass is a poor scheme choice
since in QCD its definition is ambiguous by an amount of ${\cal O}(\Lambda_{\rm
  QCD})$ (in perturbation theory this is referred to as the infrared renormalon
problem).  Nice mass schemes avoid this problem by a suitable choice for the
$a_{nk}$'s, and are known as short-distance mass schemes. Using observables
expressed in terms of short-distance schemes the accuracy that the mass of a
colored particle can be measured is not limited by $\Lambda_{\rm QCD}$. This
fact is  important to obtain the $\simeq 40\,{\rm MeV}$ uncertainties for
current measurements of the $b$-quark mass~\cite{Yao:2006px}.
%, and for the $\simeq
%100\,{\rm MeV}$ accuracy that can be achieved for the top-quark with a threshold
%scan at a linear collider~\cite{?}. 
% 
For current top-measurements based on reconstruction there is another
important restriction on viable 
short-distance schemes. In these kinematic-based analyses the top-quark
decay is treated with a Breit-Wigner in the MC's, and only
``top-resonance mass 
schemes'' with a small $R\sim \Gamma_t$ are compatible with the Breit-Wigner
form~\cite{Fleming:2007qr}. This restriction rules out directly measuring the
$\overline {\rm MS}$ mass, which has a much larger $R=m_t$.

To explore these issues in detail it is useful to consider a case with full
analytic control, namely jets produced in $e^+e^-\to t\bar t$ with c.m.\ energy
$Q \gg m_t$.  Here the jet-invariant masses $M_{t,\bar t}$ sum over particles in
the top and antitop hemisphere defined with respect to the thrust axis, and
are examples of event shape variables 
for massive quarks.  The resonance region $|M_{t,\bar t}-m_t|\ll m_t$ is most
sensitive to the top-mass. The appropriate factorization theorem was
derived in Ref.~\cite{Fleming:2007qr},
\begin{eqnarray} \label{Fthm}
 &&\hspace{-0.8cm}
 \frac{d^2\sigma}{dM_t^2 dM_{\bar t}^2} 
 = \sigma_0 H_Q(Q,\mu_m) H_m\Big(m,\frac{Q}{m},\mu_m,\mu_\Lambda\Big)
  \nonumber \\ 
 &&\hspace{-0.8cm}
 \times\!\!\int\!\! d\ell^+d\ell^- B_+(\hat s_t\! -\! \frac{Q \ell^+}{m},\delta
 m, \Gamma_{\!t},\mu_\Lambda,\mu_\Gamma) 
  \\
 && \hspace{-0.8cm}
  \times  B_-(\hat s_{\bar t}\!-\! \frac{Q\ell^-}{m},\delta m,\Gamma_{\!t},\mu_\Lambda,\mu_\Gamma) 
 S(\ell^+,\ell^-,\delta\Delta,\bar\Delta,\mu_\Lambda) , 
 \nonumber
\end{eqnarray}
where $\hat s_t \equiv (M_t^2-m^2)/m$ and $\hat s_{\bar t} \equiv (M_{\bar
  t}^2-m^2)/m$. Eq.~(\ref{Fthm}) is valid to all orders in $\alpha_s$ and has
power corrections of ${\cal O}(m\alpha_s(m)/Q, m^2/Q^2,$ $\Gamma_t/m, \hat
s_{t,\bar t}/m)$. Here $H_Q$ and $H_m$ are perturbative coefficients describing
the hard scales $Q$ and $m$, as well as the summation of large logarithms from
$Q\gg m\gg \mu_\Gamma$, where $\mu_\Gamma\sim \Gamma_t + Q\Lambda_{\rm QCD}/m+
\hat s_t$. The $B_\pm$ are perturbatively calculable heavy-quark jet functions,
which describe the evolution and subsequent decay of the top/antitop quark to
jets, and include a resummation of large logs between $\mu_\Gamma$ and the soft
scale $\mu_\Lambda$\,\raisebox{-3pt}{$\stackrel{>}{\sim}$} $\Lambda_{\rm QCD} +
 m\Gamma_t/Q+ m\hat s/Q$. A complete resummation of large logs at
next-to-leading order (NLL) was carried out in
Ref.~\cite{Fleming:2007xt}, where the reader can find further details. In
Eq.~(\ref{Fthm}) $m$ and $\delta m$ are specified in a top-resonance
mass-scheme, of which an example known as the jet-mass is discussed below in
section~2. Finally, $S$ is the hemisphere soft-function describing the
soft-radiation between jets. $S$ is perturbatively calculable for $\ell^\pm \gg
\Lambda_{\rm QCD}$ and non-perturbative for $\ell^\pm\sim \Lambda_{\rm QCD}$.
The parameter $\bar\Delta$ is a gap in the soft function~\cite{Hoang:2007vb},
while $\delta \Delta$ indicates that the scheme for this gap is defined so that
both perturbative and nonperturbative contributions are included without
inducing a leading renormalon ambiguity, as discussed below in section~3.

The result in Eq.~(\ref{Fthm}) relates an experimentally measurable hadronic
observable, $d^2\sigma/dM_t^2 dM_{\bar t}^2$,  to the short distance mass
$m$. As an example, consider the peak 
position of the invariant mass distribution $M_{t,\bar t} = M^{\rm peak}$.
Schematically this relation has the form
\begin{eqnarray} \label{Mpeak}
 M^{\rm peak} = m + \Gamma_t(\alpha_s + \alpha_s^2 + \ldots ) +
 \frac{Q\Lambda_{\rm QCD}}{m} \,,
\end{eqnarray}
where the perturbative shifts $\sim \Gamma_t(\alpha_s+\alpha_s^2+\ldots)$ can be
calculated from the jet-functions, $B_\pm$, and the nonperturbative shift $\sim
Q\Lambda_{\rm QCD}/m$ is completely determined by the soft-function $S$. Having
a systematic separation of the perturbative and non-perturbative shifts into
$B_\pm$ and $S$ is important for a precision determination of the Lagrangian
mass $m$. In particular the value of the terms in the perturbative series in
Eq.~(\ref{Mpeak}) are related to the choice of mass-scheme for $m$, and are
modified by different scheme choices. Thus, control over the mass-scheme
requires control over these corrections. Note that the non-perturbative shift is
always positive for hemisphere masses, and quite sizeable, $\simeq
1$--$2\,{\rm GeV}$ for typical values of $Q$. In sections 2 and 3 below we
discuss the functions responsible for these perturbative and nonperturbative
shifts in more detail.  Then in section 4 we quantitatively analyze the relation
between $M^{\rm peak}$ and $m$ using Eq.~(\ref{Fthm}).  Finally, in section~5 we
return to implications for the Tevatron mass, and discuss a very useful
mass-scheme for measuring both top and bottom heavy-quark masses, the
MSR-scheme.

\section{Heavy-Quark Jet Function}

Since $\hat s \ll m$ the jet function can be
formulated in the heavy-quark limit with HQET,
\begin{eqnarray} \label{Lh}
  {\cal L}_{\rm HQET} &=& \bar h_v \Big(iv\cdot D-\delta m +
  \frac{i\Gamma_t}{2}\Big) h_v \,.
\end{eqnarray}
Here $\delta m = m^{\rm pole} - m$ encodes the mass-scheme choice, and
$\Gamma_t$ is the total width of the top-quark which suffices since our observable is
inclusive in the decay products. The heavy-quark jet function is defined as the
imaginary part of a $h_v$ propagator connected by a light-like Wilson
line~\cite{Fleming:2007qr}, $B(\hat s,\delta m ,\Gamma_t,\mu)={\rm Im}[ {\cal
  B}(\hat s,\delta m,\Gamma_t,\mu)]$ where
\begin{eqnarray}
 &&\hspace{-0.7cm}
  {\cal B}(\hat s,\delta m,\Gamma_t,\mu) =
 \\
 &&\hspace{-0.7cm}
  \frac{-i}{m} \int\!\!
  \frac{d^4x}{4\pi N_c} \langle 0 | T \bar h_v(0) W_n(0) W_n^\dagger(x) h_v(x) |
  0 \rangle .\nonumber
\end{eqnarray}
At tree level in $\alpha_s$ we have a Breit-Wigner, $B=\Gamma_t/[(\pi m) (\hat
s^2+\Gamma_t^2)]$.  From Eq.~(\ref{Lh}) the jet function obeys the shift
identity ${\cal B}(\hat s,\delta m,\Gamma_t,\mu) = {\cal B}(\hat s-2 \delta m +i
\Gamma_t,0,0,\mu)\equiv {\cal B}(\hat s-2 \delta m +i \Gamma_t,\mu)$. It also
has an RGE $\mu d/d\mu {\cal B}(\hat s,\mu) = \int d\hat s' \gamma_B(\hat s-\hat
s',\mu) {\cal B}(\hat s',\mu)$ where the anomalous dimension has a plus-function
term with the universal cusp anomalous dimension and a $\delta$-function term,
$\gamma_B(\hat s,\mu) = -2 \Gamma^{\rm cusp}[\alpha_s]/\mu [ \mu\theta(\hat
s)/\hat s]_+ + \gamma_B[\alpha_s] \delta(\hat s)$. Summing logarithms from
$\mu_\Gamma$ to $\mu_\Lambda$ gives
\begin{eqnarray} \label{Brun}
 &&\hspace{-0.7cm}
 B(\hat s,\delta m, \Gamma_t,\mu_\Lambda,\mu_\Gamma)\nonumber\\
 &&\hspace{-0.7cm}
 = \int\!\! d\hat s'\,
  U_B(\hat s\! -\! \hat s',\mu_\Lambda,\mu_\Gamma) B(\hat s',\delta
  m,\Gamma_t,\mu_\Gamma) \,,  
\end{eqnarray}
which is the function $B=B_+=B_-$ appearing in Eq.~(\ref{Fthm}).
Eq.~(\ref{Brun}) sums all logs that can affect the shape of $d^2\sigma/dM_t^2
dM_{\bar t}^2$~\cite{Fleming:2007xt}. There is some freedom in combining fixed
order results for $B$ at $\mu_\Gamma$ with the RGE in $U_B$.  We define a
counting that elevates the importance of fixed order results
\begin{eqnarray}
 &&\hspace{-0.7cm}
 {\rm LL\!:} \ \ \ \ \  \mbox{$1$-loop $\Gamma_{\rm cusp}$, tree-level matching}\,;
   \\*
  &&\hspace{-0.7cm}
{\rm NLL\!:}  \ \ \: \mbox{$2$-loop $\Gamma_{\rm cusp}$, $1$-loop $\gamma_B$
   and matching} ; \nonumber \\*
  &&\hspace{-0.7cm}
{\rm NNLL\!:} \ \mbox{$3$-loop $\Gamma_{\rm cusp}$, $2$-loop $\gamma_B$
   and matching} \nonumber . 
\end{eqnarray} 
One-loop results for $B$ and $\gamma_B$ were computed in
Ref.~\cite{Fleming:2007xt}, and two-loop results in Ref.~\cite{Jain:2008gb}. The
three-loop result for $\Gamma_{\rm cusp}$ is known from Ref.~\cite{Moch:2004pa},
so $B$ is known at NNLL. Results for $B$ in moment space have also been
computed~\cite{Aglietti:2006wh,Mitov:2006xs}. A different heavy quark jet
function also occurs in the massive quark form factor, whose anomalous dimension
shares several coefficients in common with $\gamma_B$~\cite{Mitov:2006xs}.

\begin{figure}[t!]
%\vspace{9pt}
\includegraphics[width=17pc]{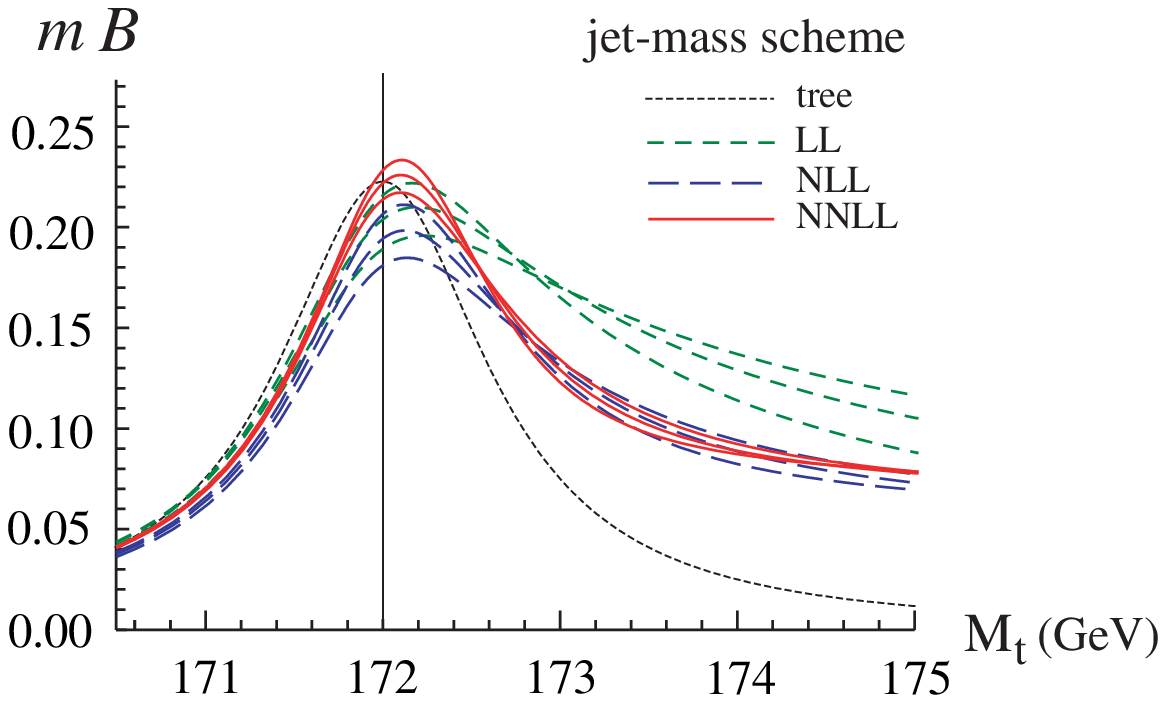}
\includegraphics[width=17pc]{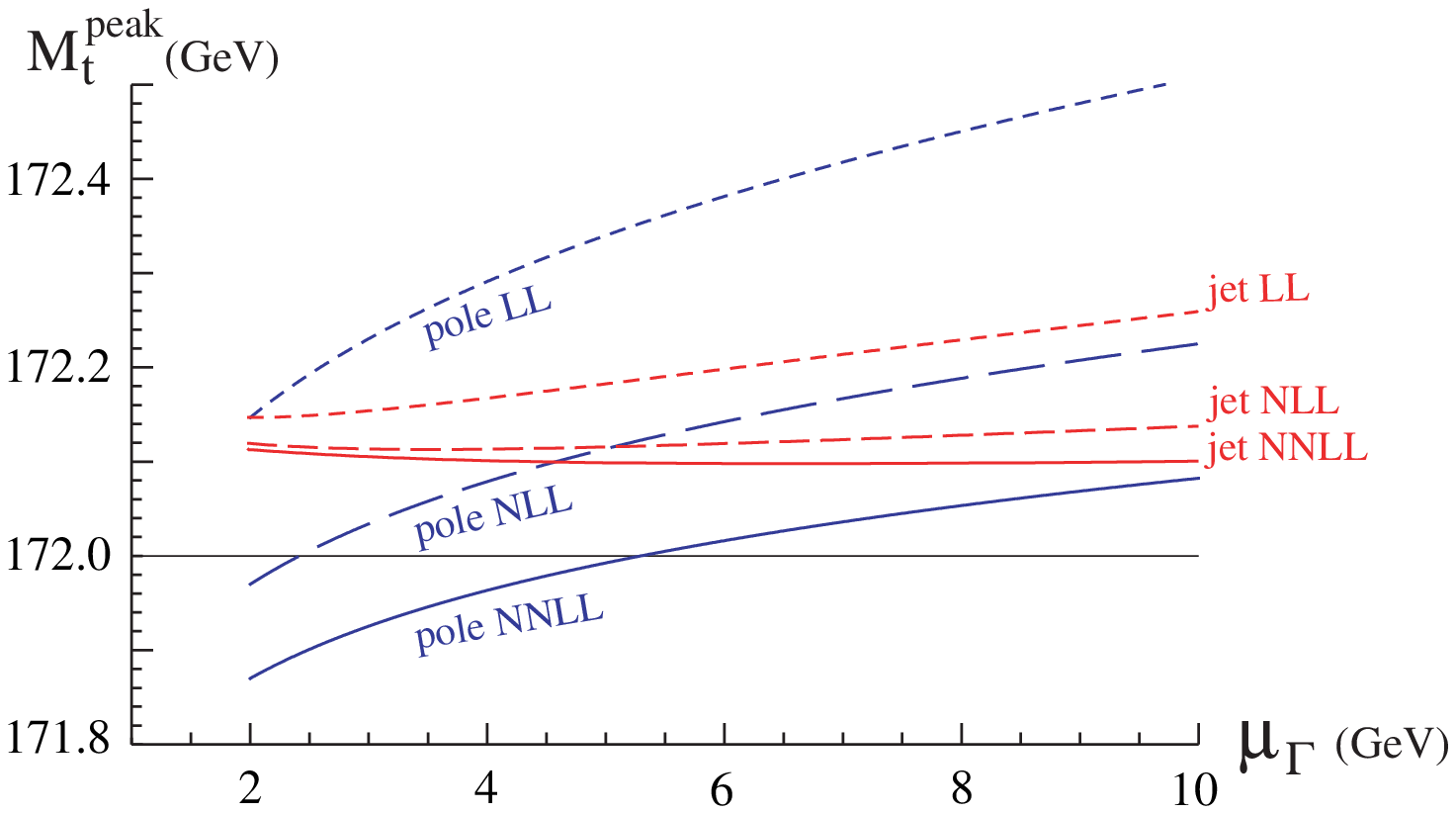}
%\framebox[55mm]{\rule[-21mm]{0mm}{43mm}}
\vspace{-0.9cm}
\caption{Heavy quark jet-function up to NNLL order (top). 
  Peak positions in the pole and jet-mass schemes (bottom). Plots
  from~\cite{Jain:2008gb}.}
\label{fig:B}
\vspace{-0.5cm}
\end{figure}
In order to implement a short-distance mass-scheme and subtract the infrared
contributions related to the pole-mass renormalon we must systematically
expand in $\delta m$ to the same level in 
$\alpha_s$ that we determine $B$ itself. Writing
\begin{eqnarray}
  \delta m =  R e^{\gamma_E} \Big\{ \frac{\alpha_s(\mu)}{\pi} \delta m_1 
   + \Big[\frac{\alpha_s(\mu)}{\pi}\Big]^2 \delta m_2 \Big\} \,,
\end{eqnarray}
with $R\sim \Gamma_t$, the remaining freedom in specifying the mass-scheme at
two-loop order corresponds to defining $\delta m_{1,2}$. The jet-function itself
can be used to define a jet-mass scheme, $m=m_J(R,\mu)$~\cite{Fleming:2007qr}.
It is convenient to use the position space jet-function $\tilde B(y,\mu)$  so
that $m_J$ has a consistent $\mu$-anomalous dimension~\cite{Jain:2008gb}, and
we define
\begin{eqnarray} \label{dmJ}
  \delta m_J \equiv \frac{R e^{\gamma_E}}{2}\: \frac{d \ln \tilde B(y,\mu)}{d\ln(i
    y)}\, \Big|_{i y e^{\gamma_E}= 1/R} \,.
\end{eqnarray}
This definition for the jet mass is possible for any value of $R$, and the
anomalous dimension in $R$ can also be consistently derived~\cite{Hoang:2008yj}.

In Fig.~\ref{fig:B} (top panel) we show results for Eq.~(\ref{Brun}) using
$R=0.8\,{\rm GeV}$, the jet-mass with reference value $m_J(R,2.0\,{\rm
  GeV})=172\,{\rm GeV}$, $\mu_\Lambda=1\,{\rm GeV}$, and three curves at each of
LL, NLL, and NNLL order corresponding to $\mu_\Gamma=3.3, 5.0,7.5\,{\rm GeV}$.
Convergence by NNLL is evident. In the bottom panel of Fig.~\ref{fig:B} we
compare the pole and jet-mass schemes and show that very good convergence for
the peak position is achieved in the jet-mass scheme.

\section{Hemisphere Soft Function}
 
The hemisphere soft function is defined by a matrix element of Wilson lines
\begin{eqnarray} \label{Sdef}
 && \hspace{-0.7cm}
 S_{\rm hemi}(\ell^+,\ell^-,\mu) =
  \frac{1}{N_c} \sum_{X_s} \delta(\ell^+ \!-\! k_s^{+a})
  \delta(\ell^- \!-\! k_s^{-b}) \nonumber \\
 &&\hspace{-0.7cm}
 \times \langle 0 | \overline Y_{\bar n}(0) Y_n(0) | X_s \rangle 
 \langle X_s | Y_n^\dagger(0) \overline Y_{\bar n}^\dagger(0) | 0 \rangle ,
\end{eqnarray}
where $Y_n^\dagger(0) = P \exp( ig \int_0^\infty ds\, n\!\cdot\! A(ns) )$ and
$\overline Y_{\bar n}^\dagger$ is similar but in the $\overline 3$
representation. Here $k_s^{+a}$ is the sum of all plus-momenta of particles in
$X_s$ in hemisphere $a$, and $k_s^{-b}$ is the sum of all minus-momenta for
particles in hemisphere $b$. Thus the $\delta$-functions in Eq.~(\ref{Sdef}) are
a reflection of the prescription for assigning soft-radiation to the invariant
masses $M_t$ and $M_{\bar t}$. This soft function is
universal~\cite{Fleming:2007qr}, the same function appears in the factorization
theorem for the hemisphere invariant mass distribution as well as the thrust and
heavy jet mass distributions of massless jets~\cite{Korchemsky:1999kt}.

$S(\ell^+,\ell^-,\mu)$ has both perturbative and non-perturbative components,
and a convenient way to account for this is~\cite{Hoang:2007vb}
\begin{eqnarray} \label{Sconv}
 &&\hspace{-0.7cm}
  S(\ell^+,\ell^-,\delta \Delta,\bar\Delta,\mu_\Lambda)  = \int\!\!
  d\ell^{+\prime} d\ell^{-\prime}
  \\
 &&\hspace{-0.7cm}
 \!\! \times S^{\rm part}(\ell^+\!-\!\ell^{+\prime},
  \ell^-\!-\!\ell^{-\prime}, \delta\Delta,\mu_\Lambda) S^{\rm
    mod}(\ell^{+\prime},\ell^{-\prime},\bar\Delta) \,.
   \nonumber
\end{eqnarray}
Here $S^{\rm part}$ refers to a partonic computation of Eq.~(\ref{Sdef}) and
encodes the proper $\mu_\Lambda$ dependence and large $\ell^\pm$ behavior into
$S$. $S^{\rm mod}$ is a model for the non-perturbative $\ell^\pm
\sim\Lambda_{\rm QCD}$ region of $S$. A gap
$\theta(\ell^{+\prime}-\bar\Delta)\theta(\ell^{-\prime}-\bar\Delta)$ is
contained in $S^{\rm mod}$ where $\bar\Delta$ is defined in a scheme specified
by $\delta\Delta(R',\mu)$. The partonic soft-function obeys $S^{\rm
  part}(\ell^+,\ell^-,\delta\Delta,\mu) = S^{\rm
  part}(\ell^+-\delta\Delta,\ell^--\delta\Delta,0,\mu)\equiv S^{\rm
  part}(\ell^+-\delta\Delta,\ell^--\delta\Delta,\mu)$. To avoid a ${\cal
  O}(\Lambda_{\rm QCD})$ renormalon (which has nothing to do with the ${\cal
  O}(\Lambda_{\rm QCD})$ renormalon of quark masses), we must systematically
expand in $\delta\Delta$ to the same level in $\alpha_s$ as we expand $S^{\rm
  part}$~\cite{Hoang:2007vb}. We write
\begin{eqnarray} 
  \delta \Delta =  R' e^{\gamma_E} \Big\{ \frac{\alpha_s(\mu)}{\pi} \delta \Delta_1 
    + \Big[\frac{\alpha_s(\mu)}{\pi}\Big]^2 \delta \Delta_2 \Big\} \,.
\end{eqnarray}
Removing the renormalon with $\delta \Delta$ avoids having to make large changes
to $S^{\rm mod}$ each time an additional order in $\alpha_s$ is included in
$S^{\rm part}$.  Much like Eq.~(\ref{dmJ}) we can define $\delta\Delta$ with the
position space soft function~\cite{Hoang:2008fs}, which yields a scheme for
$\bar\Delta(R,\mu)$ with a consistent RGE in $\mu$ and $R$
\begin{eqnarray} \label{dDelta}
 \delta\Delta = R' e^{\gamma_E} \frac{d\ln \tilde S_{\rm
     part}(x_1,x_2,\mu)}{d\ln(ix_1)} \Big|_{i e^{\gamma_E} x_{1,2}  =1/R'}.
\end{eqnarray}
Results for $S^{\rm part}(\ell^+,\ell^-,\delta\Delta,\mu)$ are known at one-loop
order~\cite{Fleming:2007xt} and two-loop order~\cite{Hoang:2008fs}, and for the
thrust soft-function in Refs~\cite{Schwartz:2007ib}.

\begin{figure}[t!]
%\vspace{9pt}
\includegraphics[width=15pc]{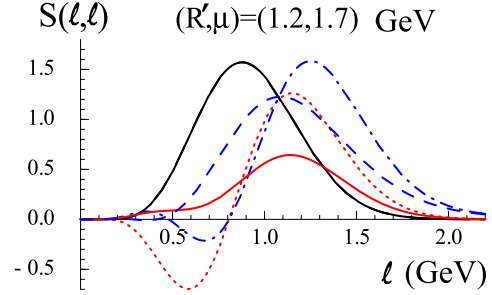}
%\framebox[55mm]{\rule[-21mm]{0mm}{43mm}}
\vspace{-0.8cm}
\caption{Diagonal hemisphere soft function up to NNLO order in the
  $\delta\Delta=0$ scheme and scheme from Eq.~(\ref{dDelta}). Plot
  from~\cite{Hoang:2008fs}.}
\label{fig:S}
\vspace{-0.5cm}
\end{figure}
In Fig.~\ref{fig:S} we compare results for Eq.~(\ref{Sconv}) at LO (black), NLO
(blue), and NNLO (red), in the $\delta\Delta=0$ scheme (dotted \& dot-dashed)
and in the $\bar\Delta$-scheme (dashed \& solid). For $S^{\rm mod}$ we use the
2d-exponential model of Ref.~\cite{Korchemsky:2000kp} with the addition of the
gap $\bar\Delta$.  The negative dips at small $\ell$ that are present in the
$\delta\Delta=0$ scheme, are removed in the renormalon free $\bar\Delta$-scheme, 
and a stable result is obtained for the peak-position of the soft function.

\section{NLL $e^+e^-\to t\bar t$ Cross-Section}

\begin{figure}[t!]
%\vspace{9pt}
\includegraphics[width=17pc]{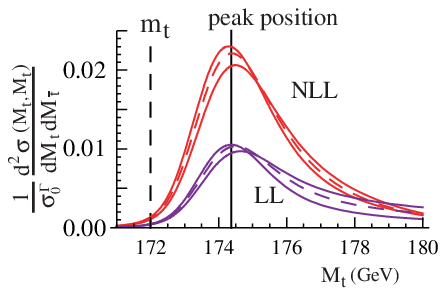}
\centerline{ \includegraphics[width=14pc]{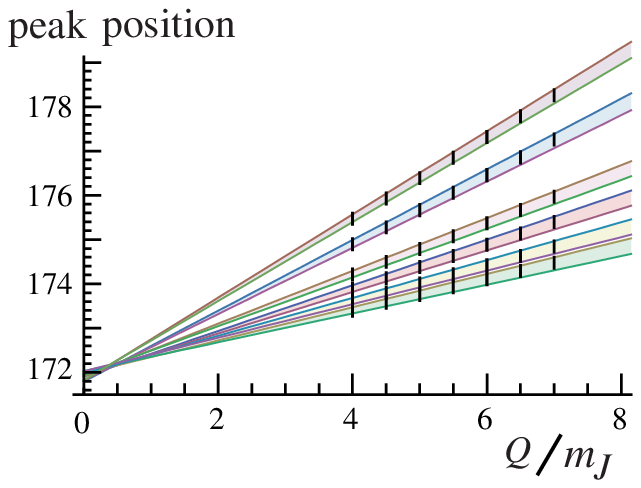} }
%\framebox[55mm]{\rule[-21mm]{0mm}{43mm}}
\vspace{-1.5cm}
\caption{Diagonal invariant mass distribution at LL and NLL order (top). Peak
  position of the NLL distribution for different soft function models versus
 $Q/m$ (bottom). Vertical black lines indicate theory uncertainties and the colored
 bands linear extrapolations to $Q/m\to 0$. 
  Plots from~\cite{Fleming:2007xt}.}
\label{fig:sig}
\vspace{-0.5cm}
\end{figure}

We now proceed to put together all the ingredients in Eq.~(\ref{Fthm}) at NLL
order~\cite{Fleming:2007xt} (since the full NNLL results have not yet been
published). The cross-section depends on $\mu_Q\sim Q$, $\mu_m\sim m$,
$\mu_\Gamma$, and $\mu_\Lambda$, and the dependence on these four
renormalization scales is reduced from LL to NLL order. The largest variation
occurs for $\mu_\Gamma$ and $\mu_\Lambda$, but is highly correlated. It becomes
significantly smaller if we vary fixing $\mu_\Gamma/\mu_\Lambda =Q/m$, which
is consistent with the scaling given below Eq.~(\ref{Fthm}).
Fig.\ref{fig:sig} (top panel) shows the invariant mass distribution for
$M_t=M_{\bar t}$ where the three curves at each order have $\mu_\Gamma=3.3, 5.0,
7.5\,{\rm GeV}$ and the same reference top-mass of $172\,{\rm GeV}$ was used as
in Fig.~\ref{fig:B}.
The ${\cal O}(\Lambda_{\rm QCD})$ renormalons associated to the pole-mass and
the partonic soft-function, both would cause bigger shifts to the location of
the invariant mass peak than the difference ($\simeq 0.15\,{\rm GeV}$) between
LL and NLL order shown in Fig.~\ref{fig:sig}.  Thus using mass and gap schemes
that avoid these renormalons was important to obtain perturbative stability. The
largest correction to $M_t^{\rm peak}-m$ is due to the soft radiation, which
causes the majority of the difference between the dashed and solid vertical
lines of Fig.\ref{fig:sig} (top panel). The advertised shift, $\propto
Q\Lambda_{\rm QCD}/m$, is demonstrated in Fig.~\ref{fig:sig} (bottom) where we
show that linear dependence on $Q$ occurs for six different models for $S^{\rm
  mod}$. For all models the extrapolations to $Q/m\to 0$ converge to the
underlying Lagrangian mass value $m$. In an $e^+e^-$ environment the correction
$M_t^{\rm peak}-m$ can be determined a) from measurements of $S$ for massless
jets that together with perturbative computations determine the shift as in
Fig.~\ref{fig:sig} (top), or b) from measurements at different $Q$'s with a
linear extrapolation that removes the soft-radiation effect as in
Fig.~\ref{fig:sig} (bottom).

\section{Implications for the Top Mass in MC programs and the
  Tevatron Top Mass}

The situation at high energy hadron colliders is somewhat different from
$e^+e^-$ collisions due to the more complicated initial state and because the
jet algorithms needed for the reconstruction of the invariant mass distribution
are implemented in terms of MC programs rather than a factorization theorem.
Nevertheless, the principles upon which final state interactions in MC's for
tops are based, have close analogues with elements in the factorization
theorem~(\ref{Fthm}). These analogies make it obvious that the top quark masses
contained in MC programs, which are measured at the Tevatron and LHC, are very
similar to the jet mass definition we discussed above. Moreover, the Tevatron
top mass measurements are statistically dominated by the reconstructed invariant
mass distribution in the peak region, which allows more reliable statements to
be made about the scheme of the Tevatron top mass.

The final state shower in MC's describes the perturbative aspects of collinear
and soft radiation of partons. Starting at transverse scales of order the
momentum transfer of the primary partons the parton shower evolves the system
down to the shower cutoff scale $R_{sc}$ which is typically in the 1~GeV range.
For the top quark the width $\Gamma_t\approx 1.5$~GeV provides an additional
natural shower cutoff for radiation off the top quark. In the factorization
theorem~(\ref{Fthm}) the same physics is described by the renormalization group
running of the factors $H_Q$ and $H_m$, the partonic contributions in the soft
function and, in particular, the jet functions $B_\pm$. As far as the question
of the mass scheme implemented in MC's is concerned, the analogue of the shower
cutoff $R_{sc}$ is the modification of the jet functions $B_\pm$ due to the
residual mass term $\delta m$, which subtracts low-energy fluctuations of the jet
functions, absorbing them into the mass definition. It is therefore the
implementation of the parton shower for the top quark and the size of the shower
cutoff $R_{sc}$ which determines the top mass definition used in the MC's.

Another important ingredient of MC's is the description of nonperturbative
effects in the hadronization process through models that depend on many
parameters. These are fixed from reference processes. For the factorization
theorem~(\ref{Fthm}) the analogue of the hadronization models is the soft
function which can be determined from event-shapes involving light quark jets.
We note that at hadron colliders the treatment of nonperturbative
effects is more involved due to the partonic initial state, and does
not necessarily lead to a positive shift of the peak position.

For hadron collisions at the Tevatron and the LHC there are
additional complications due to for example underlying events, more involved
combinatorial background and initial state showers. However, if such effects
are modeled correctly in the MC's they do not affect the correspondences 
mentioned above. We therefore conclude that 
the top quark mass that is implemented in MC's has the property
\begin{eqnarray}
\label{mcmassdef}
 \delta m_t^{\rm MC}(R_{sc}) = R_{sc} \Big[\frac{\alpha_s(\mu)}{\pi}\Big] +\ldots
\end{eqnarray}
with $R_{sc}\sim \Gamma_t\sim 1$~GeV. The exact coefficients in the series on
the RHS depend on how the parton shower is implemented and are currently
unknown.  However, it is reasonable to assume that they are of order one.
Numerically, the MC masses are therefore quite close to the jet mass scheme
defined in~(\ref{dmJ}) for which $R\sim R_{sc}$ or to any other short-distance
mass scheme that falls in the category of~(\ref{mRdef}) with $R\sim R_{sc}$.

A good way to illustrate the numerical size of the scheme uncertainty due to the
ignorance of the coefficients in~(\ref{mcmassdef}) is the associated uncertainty
in the $\overline{\mbox{MS}}$ top mass $\overline m_t(\overline m_t)$. This
uncertainty is essential for fits of electroweak precision observables and the
resulting indirect Higgs mass bounds. To avoid large logarithms in the
perturbative relation of the MC or the jet mass (with $R\sim 1$~GeV) and the
$\overline{\mbox{MS}}$ mass (with $R\sim m_t$), one can use the evolution
equation in $R$ from Ref.~\cite{Hoang:2008yj}.  Conversions involving the
$\overline{\mbox{MS}}$ mass $\overline m_t(\overline m_t)$ are particularly
simple when the R-evolution is carried out with the MSR-mass scheme $m_t^{\rm
  MSR}(R)$, which is a short-distance mass scheme that also falls in the
category of~(\ref{mRdef}) and has in addition the property $m_t^{\rm
  MSR}(m_t^{\rm MSR})=\overline m_t(\overline m_t)$.  The MSR mass definition is
obtained from the $\overline {\rm MS}$-pole series by a simple replacement
rule~\cite{Hoang:2008yj}.  We can identify the MSR mass at a low scale $R$ in
the 1~GeV range with the mass that is contained in MC programs,
\begin{eqnarray}
\label{mteverror}
m_t^{\rm MC}(R_{sc}) \, = \, m_t^{\rm MSR}(3^{+6}_{-2}~\mbox{GeV})
\,.
\end{eqnarray}
The variation of $R$ between 1 and 9~GeV parameterizes our remaining ignorance.

\begin{figure}[t!]
%\vspace{9pt}
\includegraphics[width=17pc]{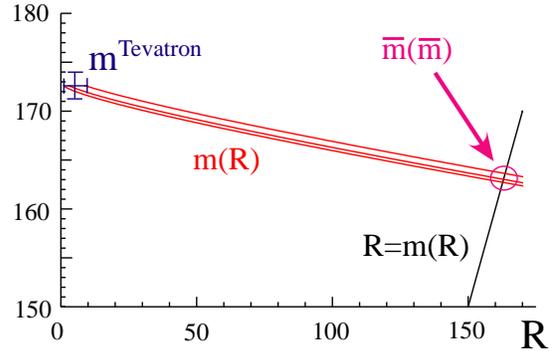}
%\framebox[55mm]{\rule[-21mm]{0mm}{43mm}}
\vspace{-1cm}
\caption{Converting the Tevatron top mass into the $\overline {\rm MS}$ scheme
  using the MSR scheme.}
\label{fig:M}
\vspace{-0.5cm}
\end{figure}

We now identify $m_t^{\rm MC}(R_{sc})$ with the Tevatron measurement $m_t^{\rm
  tev}=172.6\pm 0.8(stat) \pm 1.1(syst)\,{\rm GeV}$~\cite{Group:2008nq}.  The
result for $\overline m_t(\overline m_t)$ is illustrated in Fig.~\ref{fig:M}.
The vertical error bars on the Tevatron mass are the experimental uncertainties
and the horizontal error bars reflect the current scheme uncertainty due to
Eq.~(\ref{mteverror}).  The three red lines show the 3-loop R-evolution of the
scheme uncertainty with the MSR mass. The intersection of the red lines with the
black line that shows the equation $m_t^{\rm MSR}(R)=R$, gives the
$\overline{\mbox{MS}}$ top mass $\overline m_t(\overline m_t)=163.0\pm
1.3^{+0.6}_{-0.3}$~GeV~\cite{Hoang:2008yj}. Here the first error is the combined
experimental one, and the second error is from the scheme uncertainty in the
Tevatron mass.

The above correspondence between the shower MC and the factorization theorem is
compelling for large $p_T$ events where top quarks are fast.  A potential
concern is that tops at the Tevatron are produced with predominantly small $p_T$
and are therefore slow.  However given that the MC framework applies to this
situation, the concept of the top-mass intrinsic to the MC should be independent
of whether the MC is applied to energetic tops or to soft tops, and hence the
correspondence we have discussed should be applicable for the Tevatron as well.
Further study of this is warranted.

This work was supported in part by the Department of Energy Office of Nuclear
Science under the grant DE-FG02-94ER40818, and the EU network contract
MRTN-CT-2006-035482 (FLAVIAnet). IWS was also supported by the DOE 
%Outstanding Junior Investigator 
OJI program and Sloan Foundation.


\begin{thebibliography}{99}

\bibitem{Group:2008nq}
  T.~T.~E.~Group {\it et al.} % [CDF Collaboration],
  %``A Combination of CDF and D0 Results on the Mass of the Top Quark,''
  arXiv:0803.1683 [hep-ex].
  %%CITATION = ARXIV:0803.1683;%%

\bibitem{Borjanovic:2004ce}
  I.~Borjanovic {\it et al.},
  %``Investigation of top mass measurements with the ATLAS detector at LHC,''
  Eur.\ Phys.\ J.\  C {\bf 39S2}, 63 (2005)
  [arXiv:hep-ex/0403021].
  %%CITATION = EPHJA,C39S2,63;%%

\bibitem{Yao:2006px}
  W.~M.~Yao {\it et al.}  [Particle Data Group],
  %``Review of particle physics,''
  J.\ Phys.\ G {\bf 33}, 1 (2006).
  %%CITATION = JPHGB,G33,1;%%

\bibitem{Fleming:2007qr}
  S.~Fleming, A.~H.~Hoang, S.~Mantry and I.~W.~Stewart,
  %``Jets from Massive Unstable Particles: Top-Mass Determination,''
  Phys.\ Rev.\  D {\bf 77}, 074010 (2008)
  [arXiv:hep-ph/0703207].
  %%CITATION = PHRVA,D77,074010;%%

\bibitem{Fleming:2007xt}
  S.~Fleming, A.~H.~Hoang, S.~Mantry and I.~W.~Stewart,
  %``Top Jets in the Peak Region: Factorization Analysis with NLL Resummation,''
  Phys.\ Rev.\  D {\bf 77}, 114003 (2008)
  [arXiv:0711.2079 [hep-ph]].
  %%CITATION = PHRVA,D77,114003;%%

\bibitem{Hoang:2007vb}
  A.~H.~Hoang and I.~W.~Stewart,
  %``Designing Gapped Soft Functions for Jet Production,''
  Phys.\ Lett.\  B {\bf 660}, 483 (2008)
  [arXiv:0709.3519 [hep-ph]].
  %%CITATION = PHLTA,B660,483;%%

\bibitem{Jain:2008gb}
  A.~Jain, I.~Scimemi and I.~W.~Stewart,
  %``Two-loop Jet-Function and Jet-Mass for Top Quarks,''
  Phys.\ Rev.\  D {\bf 77}, 094008 (2008)
  [arXiv:0801.0743 [hep-ph]].
  %%CITATION = PHRVA,D77,094008;%%

\bibitem{Moch:2004pa}
  S.~Moch, J.~A.~M.~Vermaseren and A.~Vogt,
  %``The three-loop splitting functions in QCD: The non-singlet case,''
  Nucl.\ Phys.\  B {\bf 688}, 101 (2004)
  [arXiv:hep-ph/0403192].
  %%CITATION = NUPHA,B688,101;%%

\bibitem{Mitov:2006xs}
  A.~Mitov and S.~Moch,
  %``The singular behavior of massive QCD amplitudes,''
  JHEP {\bf 0705}, 001 (2007)
  [arXiv:hep-ph/0612149].
  %%CITATION = JHEPA,0705,001;%%

\bibitem{Aglietti:2006wh}
  U.~Aglietti, L.~Di Giustino, G.~Ferrera and L.~Trentadue,
  %``Resummed mass distribution for jets initiated by massive quarks,''
  Phys.\ Lett.\  B {\bf 651}, 275 (2007)
  [arXiv:hep-ph/0612073];
  %%CITATION = PHLTA,B651,275;%%
%\bibitem{Aglietti:2008xn}
%  U.~Aglietti, L.~Di Giustino, G.~Ferrera and L.~Trentadue,
  %``Comment on Resummation of Mass Distribution for Jets Initiated by Massive
  %Quarks,''
  arXiv:0804.3922 [hep-ph].
  %%CITATION = ARXIV:0804.3922;%%

\bibitem{Hoang:2008yj}
  A.~H.~Hoang, A.~Jain, I.~Scimemi and I.~W.~Stewart,
  %``Infrared Renormalization Group Flow for Heavy Quark Masses,''
  arXiv:0803.4214 [hep-ph].
  %%CITATION = ARXIV:0803.4214;%%

\bibitem{Korchemsky:1999kt}
  G.~P.~Korchemsky and G.~Sterman,
  %``Power corrections to event shapes and factorization,''
  Nucl.\ Phys.\  B {\bf 555}, 335 (1999)
  [arXiv:hep-ph/9902341];
  %%CITATION = NUPHA,B555,335;%%
%\bibitem{Bauer:2003di}
  C.~W.~Bauer, C.~Lee, A.~V.~Manohar and M.~B.~Wise,
  %``Enhanced nonperturbative effects in Z decays to hadrons,''
  Phys.\ Rev.\  D {\bf 70}, 034014 (2004)
  [arXiv:hep-ph/0309278].
  %%CITATION = PHRVA,D70,034014;%%

\bibitem{Hoang:2008fs}
  A.~H.~Hoang and S.~Kluth,
  %``Hemisphere Soft Function at O(alpha_s^2) for Dijet Production in e+e-
  %Annihilation,''
  arXiv:0806.3852 [hep-ph].
  %%CITATION = ARXIV:0806.3852;%%

\bibitem{Schwartz:2007ib}
  M.~D.~Schwartz,
  %``Resummation and NLO Matching of Event Shapes with Effective Field Theory,''
  Phys.\ Rev.\  D {\bf 77}, 014026 (2008)
  [arXiv:0709.2709 [hep-ph]];
  %%CITATION = PHRVA,D77,014026;%%
%\bibitem{Becher:2008cf}
  T.~Becher and M.~D.~Schwartz,
  %``A precise determination of alpha_s from LEP thrust data using effective
  %field theory,''
  JHEP {\bf 0807}, 034 (2008)
  [arXiv:0803.0342 [hep-ph]].
  %%CITATION = JHEPA,0807,034;%%

\bibitem{Korchemsky:2000kp}
  G.~P.~Korchemsky and S.~Tafat,
  %``On power corrections to the event shape distributions in QCD,''
  JHEP {\bf 0010}, 010 (2000)
  [arXiv:hep-ph/0007005].
  %%CITATION = JHEPA,0010,010;%%

\end{thebibliography}
\end{document}